\documentclass[aps,prb,reprint,superscriptaddress,amsmath,amssymb,showpacs]{revtex4-2}
\usepackage{braket}
\usepackage{siunitx}
\usepackage{amsmath}
\usepackage{amssymb}
\usepackage{graphicx}
\usepackage{tabularx}
\usepackage{float}

\newcommand{\RM}[1]{\MakeUppercase{\romannumeral #1}}

\begin{document}
\title{Nonadiabatic vibronic effects in single-molecule junctions: A theoretical study using the hierarchical equations of motion approach}

\author{C.\ Kaspar}
\affiliation{Institute of Physics, University of Freiburg, Hermann-Herder-Str. 3, D-79104 Freiburg, Germany}
\email{christoph.kaspar@physik.uni-freiburg.de}

\author{A.\ Erpenbeck}
\affiliation{Department of Physics, University of Michigan, Ann Arbor, Michigan 48109, USA}

\author{J.\ B\"atge}
\affiliation{Institute of Physics, University of Freiburg, Hermann-Herder-Str. 3, D-79104 Freiburg, Germany}

\author{C.\ Schinabeck}
\affiliation{Institute of Physics, University of Freiburg, Hermann-Herder-Str. 3, D-79104 Freiburg, Germany}

\author{M.\ Thoss}
\affiliation{Institute of Physics, University of Freiburg, Hermann-Herder-Str. 3, D-79104 Freiburg, Germany}
\affiliation{EUCOR Centre for Quantum Science and Quantum Computing, University of Freiburg, Hermann-Herder-Str. 3, D-79104 Freiburg, Germany}

\begin{abstract}
The interaction between electronic and vibrational degrees of freedom is an important mechanism in nonequilibrium charge transport through molecular nanojunctions. 
While adiabatic polaron-type coupling has been studied in great detail, new transport phenomena arise for nonadiabatic coupling scenarios corresponding to a breakdown of the Born-Oppenheimer approximation. 
Employing the numerically exact hierarchical equations of motion approach, we analyze the effect of nonadiabatic electronic-vibrational coupling on electron transport in molecular junctions considering a series of models with increasing complexity.
The results reveal a significant influence of nonadiabatic coupling on the transport characteristics and a variety of interesting effects, including negative differential conductance.
The underlying mechanisms are analyzed in detail.
\end{abstract}

\maketitle

\section{Introduction}
Inspired by the work of Aviram and Ratner \cite{AVIRAM1974277} charge transport through single-molecule junctions has been the focus of many experimental and theoretical studies \cite{doi101146annurevphyschem521681,nitzan2003electron,cuniberti2010introducing,cuevas2010molecular,galperin2007molecular,ZIMBOVSKAYA20111,httpsdoiorg101002pssb201350048,baldea2016molecular,doi10106315003306}.
Due to their small mass and size, charge transport in molecular nanojunctions is often strongly affected by the interaction between electronic and vibrational degrees of freedom resulting in a variety of interesting phenomena such as avalanche-like transport \cite{PhysRevB74205438,PhysRevB90075409,PhysRevB101075422,PhysRevB84205450}, current-induced dissociation \cite{PhysRevB97235452,PhysRevB102195421,doi10106350053828}, dynamical symmetry breaking \cite{PhysRevLett97166801}, local heating and cooling \cite{PhysRevLett95155505,PhysRevB83115414,PhysRevB73155306,PhysRevB83125419,PhysRevB91245429,PhysRevB98081404,doi101021acsnanolett8b01127}, vibrational instabilities \cite{PhysRevB97235429,PhysRevB83125419,PhysRevB91245429}, and switching \cite{Jan_van_der_Molen_2010}.

Most theoretical studies of vibrationally coupled charge transport in molecular junctions are based on polaron-type models, which result from the state-specific dependence of the electronic energies of the molecule on the nuclear displacement, which lead to, e.g., a change of the potential energy surface of the molecular bridge upon charging the molecule \cite{galperin2007molecular,PhysRevB83115414,PhysRevB94201407}.
However, due to the dependence of the electronic states on the nuclear coordinates, the kinetic energy operator of the nuclei may cause transitions between different electronic states which can influence the transport properties profoundly.
This mechanism, which represents a breakdown of the Born-Oppenheimer approximation, is referred to as nonadiabatic electronic-vibrational coupling \cite{doihttpsdoiorg1010029780470142813ch2,doi1011425406} and has been investigated, e.g., in the context of Jahn-Teller effects in molecular junctions \cite{PhysRevB77075323,PhysRevB78233401,Reckermann2008,PhysRevB81113106,YING201433} and in STM studies of olygothiophene molecules on a Au substrate \cite{Repp2010}.
It also manifests itself in the off-resonant transport regime in structures in inelastic electron tunneling spectra (IETS) \cite{doi101021nl052224r,doi101021nn100470s,PhysRevB86155411,httpsdoiorg101002pssb201350212}.
In this paper, we study the influence of nonadiabatic electronic-vibrational coupling on charge transport in single-molecule junctions. 
Extending our previous work in Ref.\ \cite{PhysRevB91195418}, we employ the numerically exact hierarchical equations of motion (HEOM) method and consider models exhibiting conical intersections.

A variety of theoretical approaches has been introduced and utilized to study vibrationally coupled charge transport in molecular junctions.
Methods relying on perturbation theory include the master equation approach \cite{Harbola2006,PhysRevB69245302,doi10106311768154,Rainer10,PhysRevB83115414,PhysRevB83125419}, scattering theory \cite{PhysRevB70125406,Cizek2005,Seidemann2009,Toroker2007}, and nonequilibrium Green's function formalism \cite{PhysRevLett102146801,galperin2007molecular,PhysRevB73045314,PhysRevB89205129,PhysRevB91195418,doi10106315145210}.
To overcome approximative treatments, numerically exact approaches have been employed such as the multilayer multiconfiguration time-dependent Hartree method \cite{doi10106313173823,PhysRevB88045137,doi10106313660206,doi101021jp401464b,PhysRevB90115145}, iterative path integral approaches \cite{PhysRevB85121408,doi10106314808108,httpsdoiorg101002pssb201349187}, diagrammatic quantum Monte Carlo simulations \cite{PhysRevB79153302,PhysRevB81113106,PhysRevB86081412,PhysRevB91064305,PhysRevB91155306}, the numerical renormalization group technique \cite{PhysRevB96195155,PhysRevB96195156,PhysRevB87195112,Jovchev_2015}, a combination of reduced density matrix techniques and impurity solvers \cite{PhysRevB92195143}, and the HEOM method employed in the present work.

The HEOM approach [also called hierarchical quantum master equation (HQME) method] was first introduced by Tanimura and Kubo to describe the relaxation dynamics of open quantum systems \cite{doi101143JPSJ58101,doi101143JPSJ75082001}.
Later, Yan $et$ $al.$ \cite{doi10106312938087,Zheng2012,Li2012,Zheng2009,Zheng2013,Cheng2015,doi101002wcms1269} and H{\"a}rtle $et$ $al.$ \cite{PhysRevB88235426,PhysRevB90245426,101103PhysRevB92085430,PhysRevB94121303,PhysRevB94201407} extended the formalism to investigate charge transport through nanosystems. 
The HEOM method provides an equation of motion for the reduced density matrix of an open quantum system by introducing a set of hierarchically coupled auxiliary density operators. 
It generalizes perturbative quantum master equation methods by including higher-order contributions as well as non-Markovian memory and thus allows for the systematic convergence of the results \cite{Zheng2012,doi10106350011599}.
In this paper, we utilize the HEOM formalism to investigate nonadiabatic electronic-vibrational effects in charge transport through molecular junctions.

This paper is organized as follows: We introduce the theoretical methodology in Sec.\ \ref{sec: Theory}, including a description of the model system and a brief review of the HEOM approach. 
In Sec.\ \ref{sec: Results}, we analyze nonadiabatic effects in charge transport considering a variety of models with increasing complexity.
Section \ref{sec: Conclusion} summarizes our investigations.

\section{Theory} \label{sec: Theory}
\subsection{Model system} \label{sec: Model}
To investigate nonadiabatic effects in charge transport, we consider a nanojunction, which consists of a molecule coupled to two macroscopic electrodes \cite{PhysRevB73045314,PhysRevB69245302,PhysRevB70125406,PhysRevLett95146806}.
The Hamiltonian of the composite system is given by
\begin{align}
 H =  & H_{\textrm{S}} + H_{\textrm{B}} + H_{\textrm{SB}}. \label{eq: totalH}
\end{align}
The molecule is described by a model which includes the relevant electronic states and vibrational modes,
\begin{align} \label{eq: SystemHamiltonian}
  H_{\textrm{S}}  = & \sum_{m=1}^{N_{\textrm{el}}} \epsilon^{\phantom{\dagger}}_m d^{\dagger}_m d^{\phantom{\dagger}}_m + U \sum_{m > n} d^{\dagger}_m d^{\phantom{\dagger}}_m d^{\dagger}_n d^{\phantom{\dagger}}_n \nonumber \\
  & + \sum_{\alpha=1}^{N_{\textrm{vib}}} \Omega^{\phantom{\dagger}}_{\alpha} a^{\dagger}_{\alpha} a^{\phantom{\dagger}}_{\alpha} + \sum_{\alpha,m,n} \lambda^{(\alpha)}_{mn} d^{\dagger}_m d^{\phantom{\dagger}}_n (a^{\phantom{\dagger}}_{\alpha} + a^{\dagger}_{\alpha}),
\end{align}
where we have employed units with $\hbar = 1$. 
Electrons with energy $\epsilon^{\phantom{\dagger}}_m$ are annihilated/created by the operators $d^{\phantom{\dagger}}_m$/$d^{\dagger}_m$ and the parameter $U$ characterizes the Coulomb interaction strength.
The vibrational degrees of freedom are modeled as harmonic oscillators of frequencies $\Omega_{\alpha}$, where we consider for mode $\alpha$ a set of $N_{\textrm{V},\alpha}$ vibrational basis states in the numerical calculations.
The annihilation/creation operators of vibrational mode $\alpha$ are $a^{\phantom{\dagger}}_{\alpha}$/$a^{\dagger}_{\alpha}$.

The coupling between electronic and vibrational degrees of freedom is encoded in the last term of the system Hamiltonian in Eq.\ (\ref{eq: SystemHamiltonian}), where the parameter $\lambda^{(\alpha)}_{mn}$ characterizes the interaction strength.
As discussed in Ref.\ \cite{PhysRevB91195418}, the scenario of an electronic-vibrational coupling, which is diagonal in the electronic subspace (i.e., $n=m$), corresponds to a treatment within the adiabatic or Born-Oppenheimer approximation \cite{doi1011425406,Yarkony2012,PhysRevB91195418}.
In the limit of vanishing molecule-lead coupling, a system Hamiltonian $H_{\textrm{S}}$ incorporating purely adiabatic coupling can be diagonalized analytically using the small Polaron transformation \cite{mahan2013many}.
The transport characteristics can then be rationalized by the Franck-Condon principle \cite{TF9262100536,PhysRev281182}.
Additional terms in the electronic-vibrational coupling, which are nondiagonal in the electronic subspace (i.e., $n \neq m$), enable nonadiabatic processes.
As a consequence, the Born-Oppenheimer approximation breaks down and the Franck-Condon principle fails to describe the transport behavior.
In this work, we investigate the effect of this so-called nonadiabatic coupling on charge transport in a molecular junction and its interplay with the adiabatic coupling.

The two electrodes are described by a noninteracting Fermi gas,
\begin{align}
 H_{\textrm{B}}  = &\sum_{k \in \textrm{L/R}} \epsilon^{\phantom{\dagger}}_k c^{\dagger}_k c^{\phantom{\dagger}}_k,
\end{align}
where the operators $c^{\phantom{\dagger}}_k$/$c^{\dagger}_k$ annihilate/create an electron in lead $ \textrm{L/R}$ with energy $\epsilon_k$.
For the sake of simplicity, the Fermi level of both electrodes is assumed to be $\epsilon_{\textrm{F}}=0$.

The interactions between the fermionic environment and the molecule are incorporated in $H_{\textrm{SB}}$ in Eq.\ (\ref{eq: totalH}).
The continuum of states in the leads is coupled to the $m$th molecular electronic state,
\begin{align} \label{eq: tunnelingH}
 H_{\textrm{SB}} = &\sum^{N_{\textrm{el}}}_{k \in \textrm{L/R} ,m=1} (V^{\phantom{\dagger}}_{k,m} c^{\dagger}_k d^{\phantom{\dagger}}_m + {\rm H.c.}),
\end{align}
where $V_{k,m}$ denotes the coupling strength. 
The molecule-lead coupling is characterized by the level-width function of lead $K \in \textrm{L/R}$
\begin{align}
 \Gamma^{\phantom{\dagger}}_{K,mn}(\epsilon)=2 \pi \sum_{k \in K} V^{*}_{k,m} V^{\phantom{*}}_{k,n} \delta(\epsilon - \epsilon^{\phantom{*}}_k).
\end{align}

Within the HEOM approach, we describe the system-bath interaction by introducing the bath coupling operators in a bath-interaction picture,
\begin{align}
 F^{\sigma}_{K,m} (t) = e^{i H_{\textrm{B}} t} \big( \sum_{k \in K} V^{\bar{\sigma}}_{k,m} c^{\sigma}_k \big) e^{-i H_{\textrm{B}} t},
\end{align}
with $\sigma=\pm$, $\bar{\sigma} \equiv -\sigma$, $c^{-(+)}_k \equiv c^{(\dagger)}_k$, and $V^{-(+)}_{k,m} \equiv V^{(*)}_{k,m}$. 
For system-bath couplings given by $H^{\phantom{\dagger}}_{\textrm{SB}}$ in Eq.\ (\ref{eq: tunnelingH}), the effect of the environment on the molecule can be entirely described by the two-time correlation functions of the free bath,
\begin{align}
 C^{\sigma}_{K,mn}(t-\tau) = & \braket{ F^{\sigma}_{K,m}(t) F^{\bar{\sigma}}_{K,n}(\tau) }_{\mathrm{B}}.
\end{align}
Here, we have taken the trace over bath degrees of freedom.
To obtain a closed set of equations within the HEOM framework, we expand the correlation functions by a series of exponential functions \cite{doi10106312938087,Zheng2012},
\begin{align} \label{eq: Correlationfunctionapprox}
 C^{\sigma}_{K,mn}(t-\tau) \approx V^{*}_{K,m} V^{\phantom{*}}_{K,n} \sum_{l=0}^{N_{\textrm{P}}} \eta^{\phantom{*}}_{K,\sigma,l} e^{-\gamma_{K,\sigma,l} (t-\tau)},
\end{align}
with $N_{\rm P}$ as the number of Pad\'{e} poles.
Thereby, we have chosen a constant molecule-lead coupling, $V_{k,n}=V_{K,m}$ with $k \in K$.
The Fermi distribution is represented by the Pad\'{e} decomposition scheme \cite{doiorg10106313077918,10106313484491,10106313602466} and the level-width function is assumed to be a Lorentzian,
\begin{align} \label{eq: Lorentzian}
 \Gamma_{K,mn}(\epsilon) = 2 \pi \frac{V^{*}_{K,m} V^{\phantom{2}}_{K,n} W^2_K}{(\epsilon-\mu^{\phantom{2}}_K)^2 + W^2_K}.
\end{align}
It has a single peak centered around the chemical potential $\mu_{K}$ with parameter $W_K$ as the band-width.
We consider the bias voltage drop at the contacts to be symmetric, i.e., $\mu_{\textrm{L}}=\Phi/2$ and $\mu_{\textrm{R}}=-\Phi/2$, where $\Phi$ denotes the bias voltage.
To avoid band edge effects in the transport properties, we set the band-width to $W_{K}=10^4\SI{}{\electronvolt}$, which effectively describes the wide-band limit. 
Throughout this paper, we assume symmetric molecule-lead coupling scenarios, i.e., $V_{K,m}=V \in \mathbb{R}$ and $\Gamma_{K,mn}=\Gamma$.

\subsection{Hierarchical equation of motion approach} \label{sec: HEOM}
We describe nonequilibrium charge transport with the HEOM formalism (also referred to as HQME method in this context).
We omit a detailed derivation and closely follow Refs.\ \cite{doi10106312938087,PhysRevB88235426,PhysRevB94201407,PhysRevB101075422}.
The HEOM approach provides an equation of motion for the reduced density operator by introducing a set of hierarchically coupled auxiliary density operators.
The equation of motion for the $n$th tier auxiliary density operator reads
\begin{align} \label{eq: HEOM}
 \dot{\rho}^{(n)}_{\textbf{j}} = & -i [H^{\phantom{\dagger}}_{\textrm{S}},\rho^{(n)}_{\textbf{j}}] - \sum^n_{k=1}\gamma^{\phantom{\dagger}}_{j_k}\rho^{(n)}_{\textbf{j}} \nonumber \\
  &- i \sum_{j_k} \mathcal{A}_{j_k} \rho^{(n+1)}_{\textbf{j}^+_k} -i \sum^{n}_{k=1} \mathcal{C}^{\phantom{\dagger}}_{j_k} \rho^{(n-1)}_{\textbf{j}^-_k}.
\end{align}
Here, $\rho^{(n=0)} \equiv \rho$ denotes the reduced density operator and $\rho^{(n>0)}_{\textbf{j}}$ are the auxiliary density operators, for which we have defined the multi-index vectors $\textbf{j}=j_n \ldots j_1$, $\textbf{j}^+_k=j_kj_n \ldots j_1$ and $\textbf{j}^-_k=j_n \ldots j_{k+1} j_{k-1} \ldots j_1$ with $j=(K,\sigma,l)$.
The operators $\mathcal{A}$ and $\mathcal{C}$ specify the coupling between different tiers,
\begin{subequations}\label{eq: HEOMOp}
\begin{align}
  \mathcal{A}_{j_k} \rho^{(n+1)}_{\textbf{j}^+_k} = & \sum_m V^{\phantom{\dagger}}_{K,m} ( d^{\bar{\sigma}}_m \rho^{(n+1)}_{\textbf{j}^+_k} + (-)^{n} \rho^{(n+1)}_{\textbf{j}^+_k} d^{\bar{\sigma}}_m ) ,\\
  \mathcal{C}^{\phantom{\dagger}}_{j_k} \rho^{(n-1)}_{\textbf{j}^-_k} = & \sum_m V^{\phantom{\dagger}}_{K,m} ( \eta^{\phantom{\dagger}}_j d^{\sigma}_m \rho^{(n-1)}_{\textbf{j}^-_k} + (-)^{n} \eta^{*}_j \rho^{(n-1)}_{\textbf{j}^-_k} d^{\sigma}_m ).
\end{align}
\end{subequations}
The auxiliary density operators contain information on bath-related observables such as the electrical current for lead $K$
\begin{align}
 I_K=ie \sum_{m,l} V^{\phantom{\dagger}}_{K,m} \textrm{Tr}_{\textrm{S}} (d^{\phantom{\dagger}}_m \rho^{(1)}_{(K,+,l)} - {\rm H.c.} ),
\end{align}
where $\textrm{Tr}_{\textrm{S}}$ denotes the trace over system degrees of freedom.

For the characterization of the transport behavior, the steady-state of the hierarchical equations in Eq.\ (\ref{eq: HEOM}) is of primary interest.
We utilize our recently proposed iterative solving technique to calculate the steady-state \cite{doi101021acsjpca1c02863}.

\section{Results} \label{sec: Results}
In this section, we investigate nonadiabatic vibronic effects on charge transport through single-molecule junctions. 
To this end, we employ models for a molecular junction, where the molecule is described by the system Hamiltonian in Eq.\ (\ref{eq: SystemHamiltonian}) for $N_{\textrm{el}}=2$ and $N_{\textrm{vib}} \le 2$.

To facilitate a detailed analysis, we first introduce in Sec.\ \ref{sec: IOP} a scheme to identify transport channels in the context of vibrationally coupled charge transport.
In Sec.\ \ref{sec: singlevib}, we investigate model systems with a single vibrational mode and Sec.\ \ref{sec: twovib} considers multimode systems.

\subsection{Analysis of transport processes} \label{sec: IOP}
Charge transport in models which incorporate solely adiabatic electronic-vibrational coupling can be characterized based on the Franck-Condon matrix elements \cite{mahan2013many}.
They provide the transition probability between two vibrational states in a charging or decharging process. 
We extend them to the scenario of nonadiabatic electronic-vibrational coupling by calculating the expression
\begin{align} \label{eq: FCME}
 P^{i,\sigma}_{\boldsymbol{n},\boldsymbol{\nu};\boldsymbol{\nu^\prime}} = & \sum_{\boldsymbol{n^\prime}}| \bra{\boldsymbol{n};\boldsymbol{\nu}}\chi^{\dagger} d^{\sigma}_i \chi \ket{\boldsymbol{n^\prime};\boldsymbol{\nu^\prime}} |^2.
\end{align}
Here, $\chi$ corresponds to the unitary transformation which diagonalizes the system Hamiltonian in Eq.\ (\ref{eq: SystemHamiltonian}).
We evaluate the expression in a basis of product states $\ket{\boldsymbol{n};\boldsymbol{\nu}} \equiv \ket{\boldsymbol{n}}\ket{\boldsymbol{\nu}}$, which span the subspace of both the electronic and vibrational degrees of freedom.
The electronic basis functions $\ket{\boldsymbol{n}}$ are given in the occupation number representation, i.e., $\ket{\boldsymbol{n}}=\ket{n_1 n_2}$, where $n_i \in \{0,1\}$ for the $i$th electronic level.
The vibrational basis functions $\ket{\boldsymbol{\nu}}$ correspond to the harmonic oscillator basis functions expressed in angular momentum or occupation number representation, depending on the investigated model system (see below).
More specifically, we denote with $\ket{\boldsymbol{\nu}}=\ket{\nu}$ the basis functions for a model with a single vibrational degree of freedom and consider with $\ket{\boldsymbol{\nu}}=\ket{\nu_1 \nu_2}$ the two-mode scenario.
Being initially in state $\ket{\boldsymbol{n};\boldsymbol{\nu}}$, $P^{i,\sigma}_{\boldsymbol{n},\boldsymbol{\nu};\boldsymbol{\nu^\prime}}$ in Eq.\ (\ref{eq: FCME}) provides the transition probability of populating ($\sigma = +$) or depopulating ($\sigma = -$) the $i$th noninteracting molecular electronic level with final vibrational state $\ket{\boldsymbol{\nu^\prime}}$.
Generally, the nonadiabatic transition matrix elements in Eq.\ (\ref{eq: FCME}) do not provide a complete picture of charge transport, since processes such as the nonequilibrium excitation of the vibrational mode are not considered \cite{PhysRevB83115414,PhysRevB94201407}.
Nevertheless, they provide a helpful tool to identify the dominant transport channels in vibrationally coupled charge transport.

In the sequential electron tunneling regime, transport mechanisms can be divided into several subprocesses.
To facilitate the identification of the processes, we denote in the following with $\ket{\overline{\boldsymbol{n};\boldsymbol{\nu}}}$ the eigenstate of $H_{\textrm{S}}$ which has the largest overlap with the noninteracting state $\ket{\boldsymbol{n};\boldsymbol{\nu}}$.
Being initially in the vibronic eigenstate $\ket{\overline{\boldsymbol{n};\boldsymbol{\nu}}}$ with energy $E$, the molecule is occupied by an electron transferred from the left lead resulting in the change of the molecular state to the eigenstate $\ket{\overline{\boldsymbol{n^\prime};\boldsymbol{\nu^\prime}}}$ with energy $E^\prime$.
The transport process ends with the transfer of the electron toward the right lead.
To provide an instructive interpretation, we introduce the notation $\ket{\overline{\boldsymbol{n};\boldsymbol{\nu}}} \stackrel{\text{LS}/\text{SR}}{\longrightarrow} \ket{\overline{\boldsymbol{n^\prime};\boldsymbol{\nu^\prime}}}$, which denotes the transition between the eigenstates $\ket{\overline{\boldsymbol{n};\boldsymbol{\nu}}}$ and $\ket{\overline{\boldsymbol{n^\prime};\boldsymbol{\nu^\prime}}}$ within a transfer event from the left lead L to the system S or from the system S to the right lead R.
The threshold bias voltage corresponding to the onset of the transport process is given by
\begin{align} \label{eq: energydifference}
 \Phi = 2(E^\prime - E).
\end{align}
To identify transport mechanisms, the eigenenergies of the isolated molecule are related to features in the current.

\subsection{Single-mode model systems with avoided level crossing} \label{sec: singlevib}
\begin{figure}
 \centering
 \includegraphics[width=0.5\textwidth]{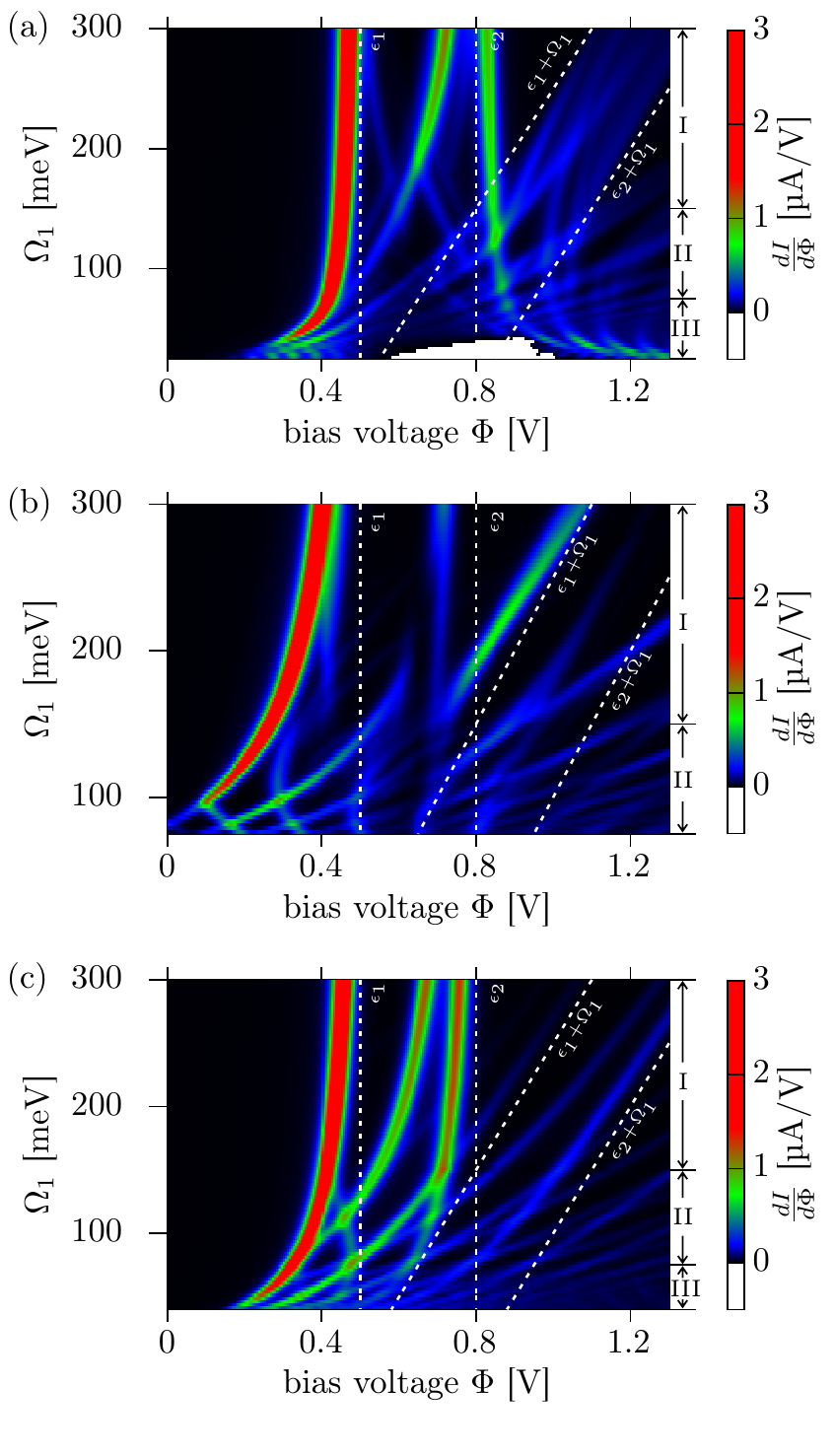}
 \caption{Conductance $\frac{d I}{d \Phi}$ as a function of bias voltage and vibrational frequency for models NONAD, INTPLY and AD (from top to bottom).
          Table \ref{tab: Models1} summarizes the parameters used.
          Furthermore, we set the temperature to $T=\SI{50}{\kelvin}$ and molecule-lead coupling strength $\Gamma=\SI{1}{\milli\electronvolt}$ to obtain sharp features in the conductance.
          We assume the Coulomb interaction strength to be $U=0$. 
          The white dashed lines indicate the positions of the unperturbed resonances.
          Three regimes can be identified: {\MakeUppercase{\romannumeral 1}} corresponds to weak, \RM{2} to intermediate, and \RM{3} to strong coupling.
          }
 \label{fig: SVF}
\end{figure}

In the scenario of a single vibrational degree of freedom, the avoided level crossing has been identified as a signature of nonadiabatic electronic-vibrational coupling in previous studies \cite{PhysRevB91195418,Repp2010}.
We investigate such systems with the two models NONAD and INTPLY, whose parameters are summarized in Table \ref{tab: Models1}.
\begin{table}[tb]
  \begin{center}
 \begin{tabularx}{\columnwidth}{XXXXXX}
 \hline
 \hline
 Model   & $\epsilon_1$ & $\epsilon_2$ & $\lambda^{(1)}_{11}$ & $\lambda^{(1)}_{22}$ & $\lambda^{(1)}_{12}$ \\
 \hline
 NONAD  & $250$ & $400$ & $0.0$ & $0.0$ & $80$  \\
 INTPLY & $250$ & $400$ & $80$  & $80$  & $80$  \\
 AD     & $250$ & $400$ & $80$  & $80$  & $0.0$ \\
 \hline
 \hline
 \end{tabularx}
\end{center}
 \caption{Overview of the model systems examined in Sec.\ \ref{sec: singlevib}, which include a single vibrational mode. 
          All parameters are given in $\SI{}{\milli\electronvolt}$.}
 \label{tab: Models1}
\end{table}
For comparison, we also show the transport properties of model AD, which solely incorporates adiabatic electronic-vibrational coupling.
In order to obtain converged results for these model systems, it is sufficient to employ second tier calculations and include $N_{\rm P}=27$ Pad\'{e} poles and up to $N_{\textrm{V},1}=35$ vibrational basis states, depending on the electronic-vibrational coupling strength.
Figure \ref{fig: SVF} depicts the conductance for the three model systems.
To identify transport mechanisms with Eq.\ (\ref{eq: energydifference}), we show in Table \ref{tab: EVSingle} in the Appendix the eigenenergies for the models NONAD and INTPLY.

We begin our investigations by introducing a simplified model for nonadiabatic charge transport and proceed with the study of model NONAD, which solely includes nonadiabatic electronic-vibrational coupling.
Finally, we examine the interplay between adiabatic and nonadiabatic coupling in model INTPLY.

\subsubsection{Effective two-state model} \label{sec: seceffmodel}
To facilitate the analysis of nonadiabatic transport properties, we define an effective (mean-field) purely electronic model with system Hamiltonian 
\begin{align}
  H_{\textrm{eff}}  = & \sum_{m=1}^{2} \epsilon^{\phantom{\dagger}}_m d^{\dagger}_m d^{\phantom{\dagger}}_m +  t^{\phantom{\dagger}}_{\textrm{eff}} (d^{\dagger}_1 d^{\phantom{\dagger}}_2 + d^{\dagger}_2 d^{\phantom{\dagger}}_1 ) \label{eq: MFSystemHamiltonian}, \\
  t^{\phantom{\dagger}}_{\textrm{eff}}= & \lambda^{(1)}_{12} {\textrm{Tr}}_{\textrm{S}}\big((a^{\phantom{\dagger}}_{1} + a^{\dagger}_{1}) \rho \big)\label{eq: teff}.
\end{align}
Here, we have introduced the effective interstate coupling $t_{\textrm{eff}}$.
It corresponds to the average displacement of the vibrational mode in a model system, which explicitly incorporates the vibrational degree of freedom.
The two-state Hamiltonian in Eq.\ (\ref{eq: MFSystemHamiltonian}) can be diagonalized analytically giving the eigenvalues and molecule-lead coupling strengths 
\begin{align}
 \tilde{\epsilon}_{1/2} = & \frac{\epsilon_1+\epsilon_2}{2} \mp \sqrt{ \Big(\frac{\epsilon_2-\epsilon_1}{2} \Big)^2 + t^2_{\textrm{eff}}} , \label{eq: eigenvalues} \\
 \tilde{\Gamma}_{1/2} = & 2\pi V^2 \big( \cos(\theta) \pm \sin(\theta)\big)^2,\label{eq: Gammaeff}\\
 \theta=&\frac{1}{2}\arctan{\Big(\frac{2t_{\textrm{eff}}}{\epsilon_1-\epsilon_2}\Big)}.\label{eq: thetaeq}
\end{align}

The effective two-state model is introduced to analyze the transport behavior of a system including nonadiabatic electronic-vibrational coupling.
To this end, we first determine the effective interstate coupling $t_{\textrm{eff}}$ in Eq.\ (\ref{eq: teff}) as a function of the bias voltage.
Subsequently, we compute the current-voltage characteristics for the two-state Hamiltonian in Eq.\ (\ref{eq: MFSystemHamiltonian}).
The results can be entirely rationalized by the eigenenergies in Eq.\ (\ref{eq: eigenvalues}) as well as the molecule-lead coupling strengths in Eq.\ (\ref{eq: Gammaeff}).

\subsubsection{Purely nonadiabatic electronic-vibrational coupling}
Here, we analyze the transport characteristics of the nonadiabatic model NONAD, whose conductance map is depicted in Fig.\ \ref{fig: SVF}(a).
The large number of transport features shows that the vibrational mode is highly involved in charge transport processes.
The conductance maps of the nonadiabatic and adiabatic models, NONAD and AD, differ substantially which indicates that the breakdown of the Born-Oppenheimer approximation results in new transport phenomena.
Since the occurring transport mechanisms depend on the dimensionless electronic-vibrational interaction strength $\lambda^{(1)}_{12}/\Omega_1$, we divide the subsequent analysis into four different coupling regimes.

\begin{figure}
 \centering
 \includegraphics[width=0.5\textwidth]{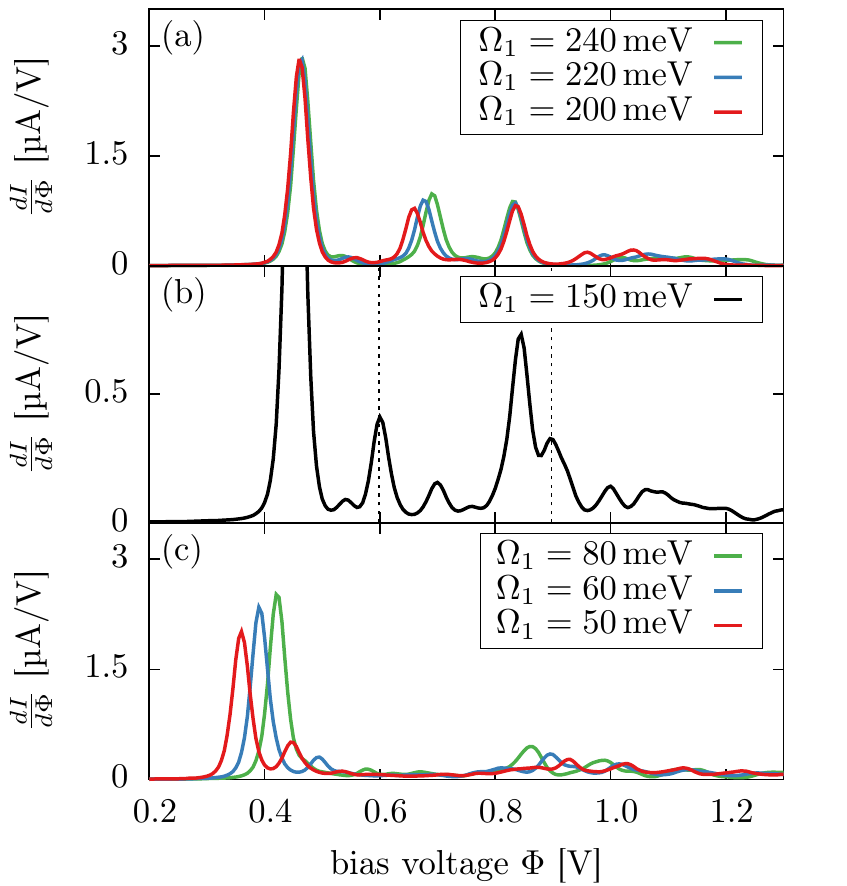}
 \caption{Conductance $\frac{d I}{d \Phi}$ as a function of bias voltage for model NONAD in the weak coupling regime in panel (a), for a frequency in resonance in panel (b), and for the intermediate coupling regime in panel (c). 
          The results correspond to a section of the conductance map in Fig.\ \ref{fig: SVF}(a) for selected vibrational frequencies and are denoted with \RM{1}, \RM{2}, and \RM{3}, respectively.
          The dashed lines indicate the onset of the transport processes $\ket{00;0} \stackrel{\text{LS}}{\longrightarrow} \ket{\overline{10;1}}$ and $\ket{00;0} \stackrel{\text{LS}}{\longrightarrow} \ket{\overline{01;0}}$.
          }
 \label{fig: OverviewNONAD}
\end{figure}

\paragraph{Weak electronic-vibrational coupling ($\lambda^{(1)}_{12} /\Omega_1 \ll 1$).}
The weak coupling regime has been previously studied in Ref.\ \cite{PhysRevB91195418} with a nonequilibrium Green's function approach, which neglects nonequilibrium vibrational processes and incorporates the electronic-vibrational coupling in a perturbative manner.
We thus expect additional transport channels to occur in the numerically exact HEOM framework used in this work.
For the sake of clarity, Fig.\ \ref{fig: OverviewNONAD}(a) depicts selected vibrational frequencies $\Omega_1 \ge \SI{200}{\milli\electronvolt}$ for regime \RM{1} in Fig.\ \ref{fig: SVF}(a).

Due to the weak electronic-vibrational coupling, the eigenenergies of the isolated molecule are close to the noninteracting energies (see Table \ref{tab: EVSingle} in the Appendix). 
The transport mechanisms are thus expected to mainly consist of purely electronic transitions as well as processes which solely involve the energetically low-lying vibrational states.
Accordingly, the three dominant features in the conductance correspond to the onset of resonant transport through the vibrational ground state of the singly and doubly charged molecule.
Besides, nonequilibrium vibrational processes, such as $\ket{00;1} \stackrel{\text{LS}}{\longrightarrow} \ket{\overline{10;1}}$ at the bias voltage $\Phi \approx \SI{0.55}{\volt}$, and transport processes, which involve higher excited vibrational states, provide a small contribution to charge transport.

\paragraph{Frequency in resonance.}
Next, we consider in Fig.\ \ref{fig: OverviewNONAD}(b) the conductance of model NONAD in a parameter regime for which the energy difference of the electronic levels equals the energy of the vibrational mode, i.e., $\epsilon_2 - \epsilon_1=  \Omega_1$.

\begin{figure*}[t]
 \includegraphics[width=0.19\textwidth]{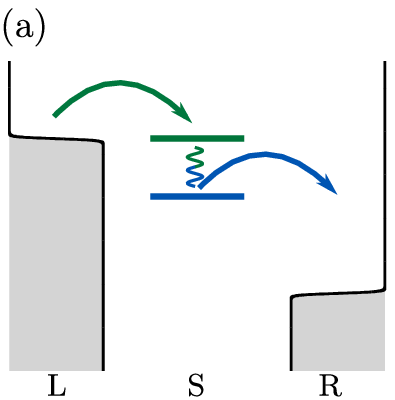}
 \includegraphics[width=0.19\textwidth]{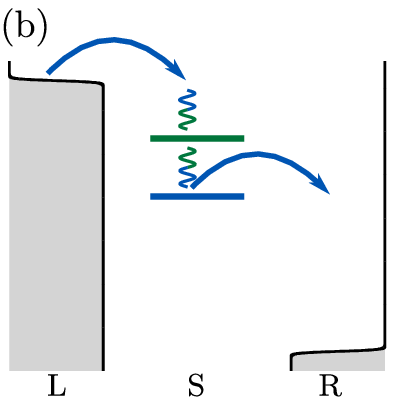}
 \includegraphics[width=0.19\textwidth]{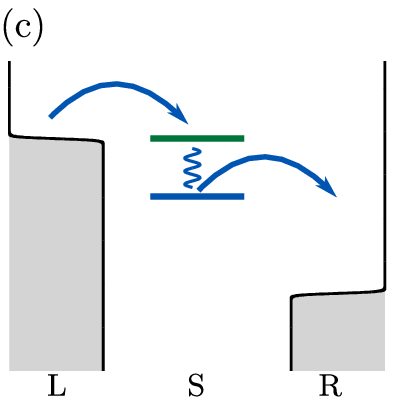}
 \includegraphics[width=0.19\textwidth]{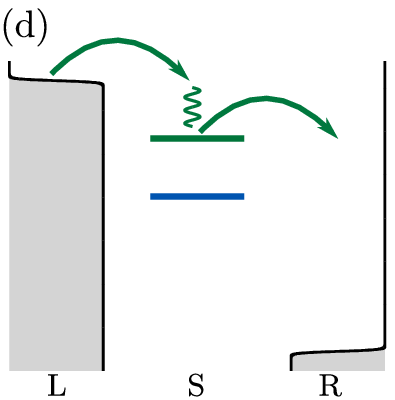}
 \includegraphics[width=0.19\textwidth]{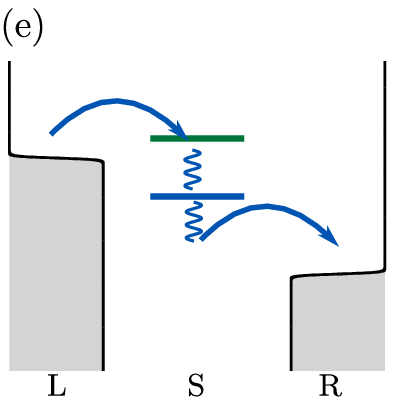}
 \caption{Schematic representation of nonadiabatic electronic-vibrational processes (a,b) and adiabatic polaron-type transport mechanisms in panels (c-e).
          The abbreviation L (R) denotes the left (right) lead and S corresponds to the system.
          Arrows indicate transport electrons and the color indicates, on which state the electrons are transferred.
         }
 \label{fig: NONADProc}
\end{figure*}

Satisfying the resonance condition enhances the influence of nonadiabatic electronic-vibrational processes on the transport properties, as shown in Refs.\ \cite{PhysRevB91195418,Repp2010}.
The processes comprise the population of one electronic level followed by an intramolecular transition to the other level.
The electron then leaves the molecule toward the right lead, as schematically shown in Fig.\ \ref{fig: NONADProc}(a).

As an example for a nonadiabatic process, we focus on the transport mechanisms occurring for the transitions $\ket{00;0} \stackrel{\text{LS}}{\longrightarrow} \ket{\overline{10;1}}$ and $\ket{00;0} \stackrel{\text{LS}}{\longrightarrow} \ket{\overline{01;0}}$.
The bias voltages at which the transport channels become active are indicated by the dashed black lines in Fig.\ \ref{fig: OverviewNONAD}(b).
We note the slightly different peak intensities of the two features in the conductance.
To characterize this nonadiabatic process, \citeauthor{Repp2010} introduced in Ref.\ \cite{Repp2010} a simplified two-state model which neglects all noninteracting states except $\ket{10;1}$ and $\ket{01;0}$.
Due to the level-mixing property of the nonadiabatic coupling, the two states would then provide equivalent transport channels in the case of the resonance condition and thus, the simplified model predicts symmetric peak intensities.
Thus, the different peak intensities indicate that processes, where more than one vibrational quanta are excited, need to be considered to capture the entire picture of the nonadiabatic process, as sketched in Fig.\ \ref{fig: NONADProc}(b).

\paragraph{Intermediate electronic-vibrational coupling ($\lambda^{(1)}_{12} / \Omega_1 \approx 1$).}
For vibrational frequencies in the range between $\Omega_1=\SI{50}{\milli\electronvolt}$ and $\Omega_1=\SI{150}{\milli\electronvolt}$, the electronic-vibrational coupling strength enters the intermediate regime, which was not covered in previous studies.
Selected vibrational frequencies for the intermediate regime \RM{2} in Fig.\ \ref{fig: SVF}(a) are depicted in Fig.\ \ref{fig: OverviewNONAD}(c).
For model AD, which incorporates purely adiabatic coupling, it marks the onset of the well-known Franck-Condon blockade \cite{PhysRevB74205438}.

\begin{figure}
 \centering
 \includegraphics[width=0.5\textwidth]{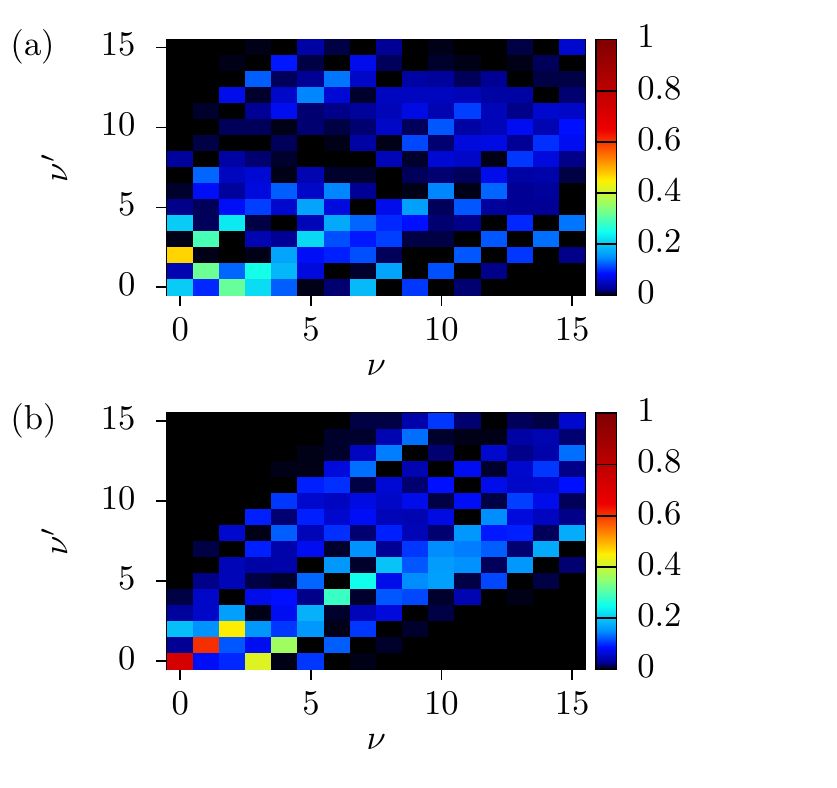}
 \caption{Nonadiabatic transition probabilities $P^{1,+}_{00,\nu;\nu^\prime}$ for $\Omega_1=\SI{50}{\milli\electronvolt}$ in panel (a) and for $\Omega_1=\SI{80}{\milli\electronvolt}$ in panel (b) for model NONAD. 
          For models with a single vibrational mode, we use the harmonic oscillator basis functions $\ket{\nu}$ in occupation number representation, where $\nu \in \mathbb{N}_0 $ corresponds to the excitation number.
          }
 \label{fig: IVFFCME}
\end{figure}

Similar to the Franck-Condon blockade, the model NONAD reveals the onset of current suppression in the low-bias voltage regime for intermediate electronic-vibrational couplings.
We investigate the phenomenon with the nonadiabatic transition probabilities $P^{1,+}_{00,\nu;\nu^\prime}$ [see Eq.\ (\ref{eq: FCME})], which are depicted in Fig.\ \ref{fig: IVFFCME} for different vibrational frequencies.
They provide the probability of charging the unoccupied molecular levels by electrons transferred from the leads. 
Thereby, the initial vibrational state $\ket{\nu}$ changes to $\ket{\nu^\prime}$.
Unlike the well-known Franck-Condon matrix elements, the transition probabilities of model systems with nonadiabatic coupling are not symmetric under the exchange of $\ket{\nu}$ with $\ket{\nu^\prime}$ due to the level-mixing property.
Transitions between energetically low-lying vibrational states become less likely with decreasing frequency and, as a consequence, the current is suppressed in the low-bias voltage regime.
This is accompanied by an enhanced influence of transport processes involving higher excited vibrational states of the lower electronic level such as the transition $\ket{00;0} \stackrel{\text{LS}}{\longrightarrow} \ket{\overline{10;2}}$ which is sketched in Fig.\ \ref{fig: NONADProc}(b).

In the voltage range between $\Phi = \SI{0.6}{\volt}$ and $\Phi = \SI{0.8}{\volt}$, transport channels with a small contribution to charge transport become active.
Further increasing the bias voltage leads to the resonant tunneling through the second electronic state including its vibrational satellites.

\paragraph{Strong electronic-vibrational coupling ($\lambda^{(1)}_{12} / \Omega_1 \gg 1$).}
Next, we consider the conductance in Fig.\ \ref{fig: SVF}(a) for frequencies $\Omega_1 \le \SI{40}{\milli\electronvolt}$ in regime \RM{3}.
For large coupling strengths, the transport characteristics exhibit negative differential conductance (NDC). 
It is noted, that nonadiabatic coupling has been reported as the source of NDC in previous investigations \cite{PhysRevB91195418}, however, in different scenarios including asymmetric molecule-lead coupling strengths, weak electronic-vibrational couplings, and quasidegenerate molecular electronic levels.

\begin{figure}
 \centering
 \includegraphics[width=0.5\textwidth]{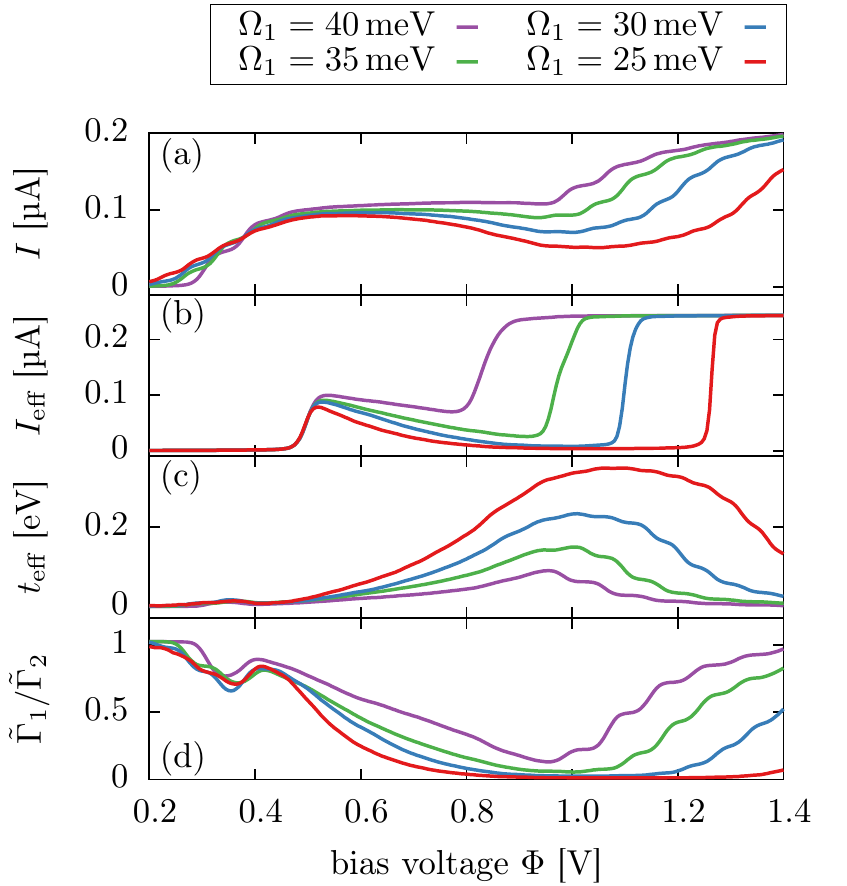}
 \caption{Current for system Hamiltonian of the complete model in Eq.\ (\ref{eq: SystemHamiltonian}), current for the effective two-state Hamiltonian in Eq.\ (\ref{eq: MFSystemHamiltonian}), effective interstate coupling in Eq.\ (\ref{eq: teff}) and ratio of the molecule-lead coupling strengths in Eq.\ (\ref{eq: Gammaeff}) (from top to bottom).}
 \label{fig: NDRExpl}
\end{figure}

To analyze the mechanism responsible for the NDC, we utilize the effective two-state model introduced in Sec.\ \ref{sec: seceffmodel}.
Figure \ref{fig: NDRExpl} depicts the current $I_{\textrm{eff}}$ of the effective model in (b), the effective interstate coupling $t_{\textrm{eff}}$ in (c) and the ratio of the molecule-lead coupling strength of the lower to the upper eigenstate in (d).
In addition, the current $I$ obtained with the system Hamiltonian of the complete model [Eq.\ (\ref{eq: SystemHamiltonian})] is shown in Fig.\ \ref{fig: NDRExpl}(a).

The discrepancy to the current of the complete model indicates that the effective two-state model cannot provide a complete picture of the transport processes.
Nonetheless, the reduced model facilitates the analysis of the NDC, as described in the following.
We first focus on the frequency $\Omega_1 = \SI{25}{\milli\electronvolt}$, since the feature is particularly dominant in this case.
NDC occurs in the bias voltage range between $\Phi\approx\SI{0.5}{\volt}$ and $\Phi\approx\SI{1.05}{\volt}$, where the upper eigenstate is located outside of the bias window and, thus, the only active transport channel is the lower eigenstate.
The effective interstate coupling increases in this regime along with the bias voltage, which results in the reduction of the effective molecule-lead coupling strength of the lower eigenstate [see Eq.\ (\ref{eq: Gammaeff})].
Hence, the current also decreases until the upper eigenstate of the two-state model enters the bias window.
For frequencies $\Omega_1 \ge \SI{30}{\milli\electronvolt}$, the effective interstate coupling is generally lower due to the smaller electronic-vibrational coupling strength.
Accordingly, the asymmetry in the molecule-lead coupling is less pronounced and thus is the NDC.

Although the molecule-lead coupling strength of the eigenstates cannot be derived analytically for the vibrationally coupled model NONAD, the effective two-state model indicates that the source of NDC in Fig.\ \ref{fig: NDRExpl}(a) is a bias-dependent asymmetry in the molecule-coupling strength induced by a strong level-mixing.
Test calculations reveal that the occurrence of NDC is robust against variations of model parameters, e.g., increasing electronic level spacing or temperature of the electrodes.
However, coupling the vibrational mode to an additional phononic environment \cite{PhysRevB103235413}, which induces vibrational relaxation, results in a less pronounced feature.

The basic requirement for the occurrence of NDC in the reduced two-state model are strong effective interstate couplings $t_{\textrm{eff}}$ which, according to Eq.\ (\ref{eq: teff}), are caused by large average displacements in the vibrationally coupled model.
First focusing on $\Omega_1 = \SI{25}{\milli\electronvolt}$, the interstate coupling in Fig.\ \ref{fig: NDRExpl}(c) increases along with the bias voltage in the range between $\Phi\approx\SI{0.4}{\volt}$ and $\Phi\approx\SI{1.05}{\volt}$. 
In this regime, the nonadiabatic process $\ket{00;0} \stackrel{\text{LS}}{\longrightarrow} \ket{\overline{10;1}}$ drives the vibrational mode toward a nonequilibrium distribution and induces large average displacements resulting in strong interstate couplings.
The decrease of $t_{\textrm{eff}}$ for bias voltages $\Phi \gtrsim \SI{1.05}{\volt}$ can be referred to the fact that the transport channel $ \ket{00;0} \stackrel{\text{LS}}{\longrightarrow} \ket{\overline{01;0}}$ becomes active which enables the transfer of electrons with a less excited vibrational mode.
Reducing the electronic-vibrational coupling strength by increasing the frequency results in a less likely occurrence of the nonadiabatic process and thus, a decreased average displacement.

\subsubsection{Interplay between adiabatic and nonadiabatic electronic-vibrational coupling}
Electronic and vibrational degrees of freedom in molecules may be subject to both adiabatic and nonadiabatic coupling.
This scenario is represented by model INTPLY with parameters summarized in Table \ref{tab: Models1}.
In the following, we investigate the transport characteristics of this model, whose conductance map is depicted in Fig.\ \ref{fig: SVF}(b), in different regimes of the dimensionless electronic-vibrational interaction strength $\lambda^{(1)}_{ij}/\Omega_1$.

\begin{figure}
 \centering
 \includegraphics[width=0.5\textwidth]{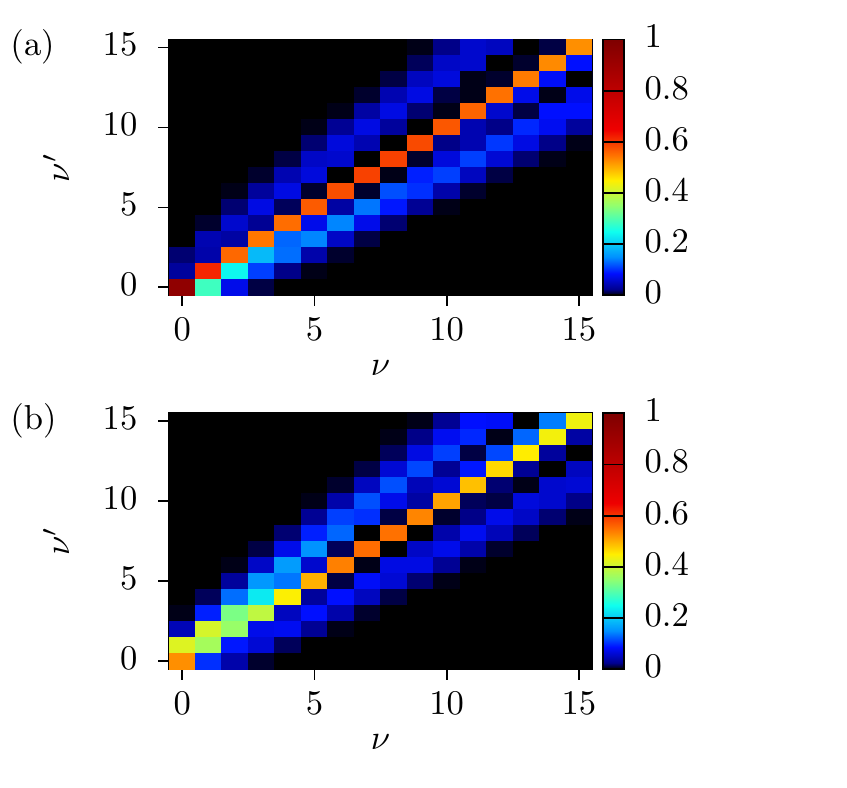}
 \caption{Nonadiabatic transition probabilities $P^{1,+}_{00,\nu;\nu^\prime}$ in panel (a) and $P^{2,+}_{00,\nu;\nu^\prime}$ in panel (b) for model INTPLY. 
          The vibrational frequency is set to $\Omega_1=\SI{240}{\milli\electronvolt}$.
          }
 \label{fig: ILVFFCME}
\end{figure}

\paragraph{Weak electronic-vibrational coupling ($\lambda^{(1)}_{ij} / \Omega_1 \ll 1$).}
The weak coupling regime of model INTPLY has also been studied in Ref.\ \cite{PhysRevB91195418} using the perturbative nonequilibrium Green's function approach. 
As shown below, there are differences to the results of the HEOM method which motivate further investigations.
To this end, Fig.\ \ref{fig: ILVFFCME} depicts the nonadiabatic transition probabilities [Eq.\ (\ref{eq: FCME})] of model INTPLY for the vibrational frequency $\Omega_1=\SI{240}{\milli\electronvolt}$.
Similar to the Franck-Condon matrix elements in an adiabatic polaron-type model, the off-diagonal transitions are clustered along a parabola.
The nonadiabatic electronic-vibrational coupling induces an asymmetry along the parabola due to the level-mixing property.
As a consequence, the charging process, which occupies the upper (lower) molecular electronic state, has an enhanced probability to (de)-excite the vibrational mode.
The interplay between adiabatic and nonadiabatic coupling additionally results in diagonal transitions $\ket{\nu}\rightarrow\ket{\nu}$.

\begin{figure}
 \centering
 \includegraphics[width=0.5\textwidth]{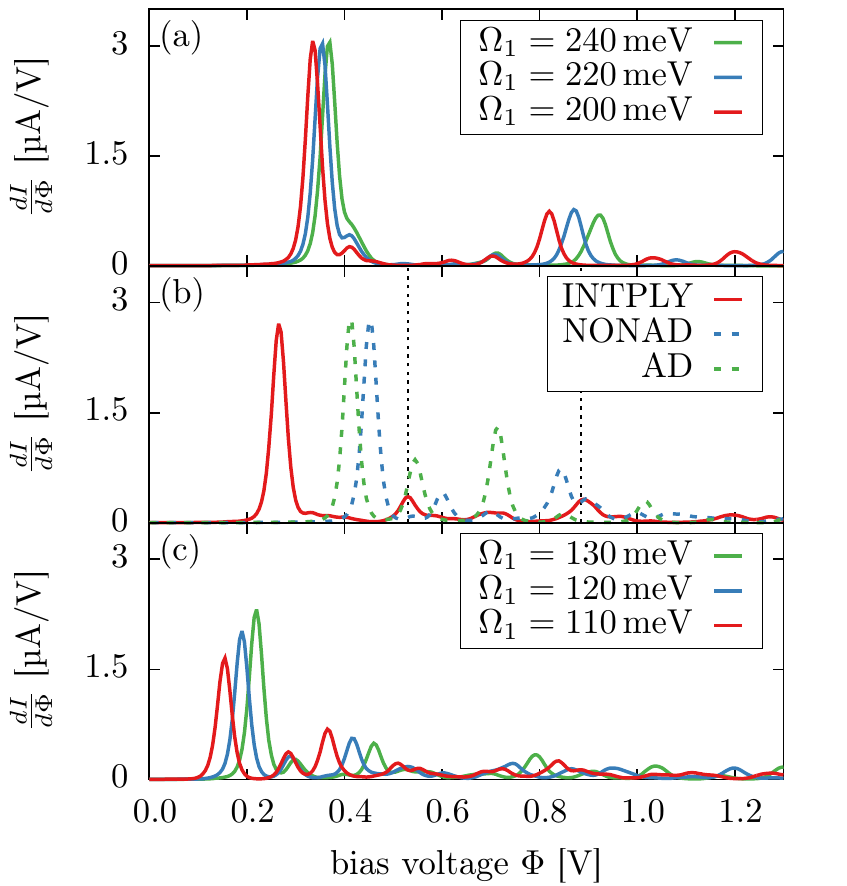}
 \caption{(a): Conductance for the weak coupling regime of model INTPLY.
          (b): Comparison of models AD, NONAD, and INTPLY for a vibrational frequency which satisfies the resonance condition $\epsilon_2 - \epsilon_1=  \Omega_1$.
          The dashed lines indicate the onset of the transport processes $\ket{00;0} \stackrel{\text{LS}}{\longrightarrow} \ket{\overline{10;1}}$ and $\ket{00;0} \stackrel{\text{LS}}{\longrightarrow} \ket{\overline{01;0}}$.
          (c): Conductance in the scenario of intermediate electronic-vibrational couplings for model INPLY.
          The results correspond to sections of the conductance maps in Fig.\ \ref{fig: SVF} for selected vibrational frequencies and are denoted with \RM{1} and \RM{2}, respectively.
         }
 \label{fig: INTPLYOverview}
\end{figure}

A section of the conductance map in Fig.\ \ref{fig: SVF}(b) is shown for selected vibrational frequencies in Fig.\ \ref{fig: INTPLYOverview}(a) for the weak coupling regime \RM{1}.
As a result of the large diagonal nonadiabatic transition probabilities, the transport mechanisms mainly consist of purely electronic transitions including the vibrational ground state of the singly and doubly charged molecule.
We note that the peak intensity corresponding to the onset of the transport process $\ket{00;0} \stackrel{\text{LS}}{\longrightarrow} \ket{\overline{10;0}}$ is significantly larger than $\ket{00;0} \stackrel{\text{LS}}{\longrightarrow} \ket{\overline{01;0}}$, which is consistent with the corresponding nonadiabatic transition probabilities.
Additional transport mechanisms with a small contribution to charge transport correspond to nonequilibrium vibrational processes, such as $\ket{00;1} \stackrel{\text{LS}}{\longrightarrow} \ket{\overline{01;0}}$ or $\ket{00;1} \stackrel{\text{LS}}{\longrightarrow} \ket{\overline{10;1}}$, which occur, for example, at the bias voltages $\Phi\approx\SI{0.41}{\volt}$ and $\Phi\approx\SI{0.62}{\volt}$ in the case of $\Omega_1 = \SI{200}{\milli\electronvolt}$.

\paragraph{Frequency in resonance.}
The influence of nonadiabatic processes on the transport properties is enhanced for model parameters which satisfy the resonance condition $\epsilon_2 - \epsilon_1=  \Omega_1$.
In Fig.\ \ref{fig: INTPLYOverview}(b), we investigate how these processes are affected by the interplay between adiabatic and nonadiabatic electronic-vibrational coupling.

Besides the different peak positions, the transport processes $\ket{00;0} \stackrel{\text{LS}}{\longrightarrow} \ket{\overline{10;1}}$ and $\ket{00;0} \stackrel{\text{LS}}{\longrightarrow} \ket{\overline{01;0}}$, which are indicated by the dashed black lines in Fig.\ \ref{fig: INTPLYOverview}(b), have peak intensities which are comparable to model NONAD.
We thus conclude that the dominant transport mechanism is the nonadiabatic process sketched in Fig.\ \ref{fig: NONADProc}(a) and adiabatic transport processes, as shown in Figs.\ \ref{fig: NONADProc}(c) and \ref{fig: NONADProc}(d), play a minor role.

Characteristic for model systems with purely adiabatic electronic-vibrational coupling is the renormalization of the electronic energies \cite{TF9262100536,PhysRev281182}, which is also referred to as the polaron shift.
The interplay between adiabatic and nonadiabatic coupling increases the renormalization profoundly for the resonance of the first electronic level.

\paragraph{Intermediate electronic-vibrational coupling ($\lambda^{(1)}_{ij} / \Omega_1 \approx 1$).}
For vibrational frequencies in the range between $\Omega_1=\SI{110}{\milli\electronvolt}$ and $\Omega_1=\SI{130}{\milli\electronvolt}$ the large deviation of the eigenenergies (see Table \ref{tab: EVSingle} in the Appendix) from the noninteracting energies results in a transport behavior which is strongly influenced by the vibrational mode.
Accordingly, the conductance depicted in Fig.\ \ref{fig: INTPLYOverview}(c), which corresponds to the regime \RM{2} in Fig.\ \ref{fig: SVF}(b), shows a large number of transport channels.

Along with increasing electronic-vibrational coupling strengths, the onset of resonant transport through the vibrational ground state of the first electronic state is shifted to lower bias voltages.
The accompanying reduction of the corresponding peak intensity marks the onset of low-bias current suppression already for $\lambda^{(1)}_{ij}/\Omega_1<1$.
The importance of vibrationally excited states is enhanced, as the increased step height of the processes $\ket{00;0} \stackrel{\text{LS}}{\longrightarrow} \ket{\overline{10;1}}$ and $\ket{00;2} \stackrel{\text{LS}}{\longrightarrow} \ket{\overline{10;0}}$ indicates.
The corresponding mechanism of the latter transport process is sketched in Fig.\ \ref{fig: NONADProc}(e), where two vibrational quanta are excited in a decharging process.

In the regime of bias voltages $\Phi \gtrsim \SI{0.9}{\volt}$, the interplay of adiabatic and nonadiabatic electronic-vibrational coupling results in the onset of transport through a large number of transport channels with a small contribution to the charge transport.

\subsection{Nonadiabatic effects in multimode systems} \label{sec: twovib}
In molecular nanojunctions with polyatomic molecules, typically more than a single vibrational degree of freedom participates in charge transport representing a significantly more complex scenario \cite{PhysRevB77075323,PhysRevB78233401,Reckermann2008,YING201433,PhysRevB90075409,PhysRevB77205314,vanderLit2013,doi101021nn405335h}.
In this section, we extend our studies to the simplest case of a two-state two-mode model, which is described by the system Hamiltonian in Eq.\ (\ref{eq: SystemHamiltonian}) for $N_{\textrm{el}}=2$ and $N_{\textrm{vib}} = 2$.
We choose the electronic-vibrational coupling to be adiabatic for the first ($\lambda=\lambda^{(1)}_{11}=-\lambda^{(1)}_{22}$) and nonadiabatic ($\lambda=\lambda^{(2)}_{12}$) for the second mode.
A large Coulomb repulsion $U$ is assumed which prevents the double charging of the molecule.

Adiabatic potential energy surfaces of polyatomic molecules can generally intersect.
At the molecular geometry, for which the surfaces become degenerate, a so-called conical intersection is formed \cite{doi1011425406}.
In the vicinity of the crossing, the vibronic coupling between the adiabatic electronic states diverges which results in the breakdown of the Born-Oppenheimer approximation.
The probability of nonadiabatic processes is significantly enhanced and thus, such model systems are well suited to investigate nonadiabatic effects in charge transport.

In the following, we study charge transport through model molecules which differ by the energy spacing of the electronic states.
We first investigate in Sec.\ \ref{sec: NonD} the influence of nonadiabatic effects on charge transport in nonsymmetric systems including nondegenerate electronic levels and proceed in Sec.\ \ref{sec: DegenerateD} with a model of degenerate electronic states exhibiting Jahn-Teller effect.

\subsubsection{Transport in a nondegenerate two-state two-mode model} \label{sec: NonD}
Here, we study nonadiabatic effects in charge transport of molecular nanojunctions with two nondegenerate noninteracting electronic levels, i.e., with energy spacings which are larger than the broadening induced by the coupling to the leads as well as the thermal broadening.
In the following, we assume the electronic-vibrational coupling strength to be $\lambda=\SI{0.08}{\electronvolt}$ and set the electronic energies to $\epsilon_1=\SI{250}{\milli\electronvolt}$ and $\epsilon_2=\SI{400}{\milli\electronvolt}$.

Important for the analysis are the eigenenergies, which are depicted in Table \ref{tab: EVTNonD} in the Appendix for selected energetically low-lying eigenstates.
We note that in two-mode models, the eigenenergies of charged vibrationally excited states cannot be related in a simple way to the fundamental frequencies $\Omega_1$ and $\Omega_2$.
While the coupling to the adiabatically coupled vibrational mode maintains the levels spacing between vibrationally excited states, the nonadiabatically coupled mode induces an enhanced level spacing.
Accordingly, the bias voltage difference between two transport channels does not correspond to multiples of the frequencies \cite{doi10108000268978100101721,PhysRevB78233401}.

\begin{figure}
 \centering
 \includegraphics[width=0.5\textwidth]{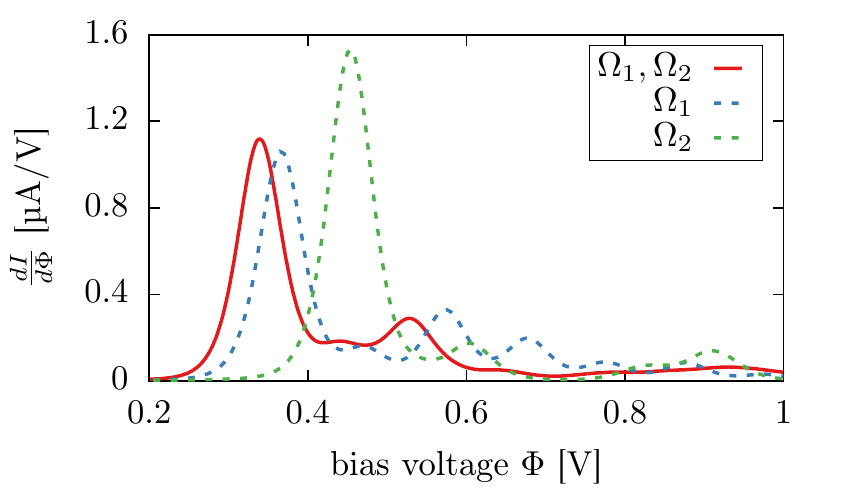}
 \caption{Conductance $\frac{d I}{d \Phi}$ as a function of bias voltage for the electronic-vibrational coupling to one or two vibrational modes with frequencies $\Omega_1=\SI{100}{\milli\electronvolt}$ and $\Omega_2=\SI{150}{\milli\electronvolt}$.
          The blue (green) dashed line corresponds to a model system incorporating a (non)adiabatically coupled single vibrational mode with a large Coulomb repulsion $U$.
          Furthermore, we set temperature $T=\SI{100}{\kelvin}$ and $\Gamma=2\pi V^2=\SI{1}{\milli\electronvolt}$.}
 \label{fig: OneModeComp}
\end{figure}

\paragraph{Comparison to single-mode models.}
We begin our analysis by comparing the transport behavior of the two-state two-mode model to systems with a single vibrational degree of freedom, to determine which transport phenomena are induced in multimode systems.
To this end, we choose the vibrational frequencies such that significant adiabatic ($\Omega_1=\SI{100}{\milli\electronvolt}$) and nonadiabatic ($\Omega_2=\SI{150}{\milli\electronvolt}$) vibrational effects are expected.
We depict in Fig.\ \ref{fig: OneModeComp} the conductance as a function of the bias voltage for a two-mode model and two models with a single vibrational degree of freedom.
The transport characteristics of similar single-mode systems has been studied in Sec.\ \ref{sec: singlevib}, however, we consider here a large Coulomb repulsion.

Although the results in Fig.\ \ref{fig: OneModeComp} show common features, the transport processes of the two-mode system cannot be entirely rationalized in terms of single-mode models.
Nevertheless, in the low-bias voltage regime, the conductance of the two-mode system and the model, which includes a single adiabatically coupled vibrational mode with frequency $\Omega_1$, are qualitatively similar.
Besides varying step heights, the main difference between the two curves are the peak positions which are located at lower bias voltages for the two-mode model.
Based on the observed similarities, we conclude that the first three transport processes in the two-mode model, which correspond to $\ket{00;00} \stackrel{\text{LS}}{\longrightarrow} \ket{\overline{10;00}}$, $\ket{00;10} \stackrel{\text{LS}}{\longrightarrow} \ket{\overline{01;00}}$ and $\ket{00;00} \stackrel{\text{LS}}{\longrightarrow} \ket{\overline{10;10}}$, are dominated by adiabatic transport mechanisms in the regime of low to intermediate bias voltages.

For bias voltages $\Phi \gtrsim \SI{0.6}{\volt}$, the results indicate that including two vibrational modes results in many transport processes with a small contribution to charge transport.

\begin{figure}
 \centering
 \includegraphics[width=0.5\textwidth]{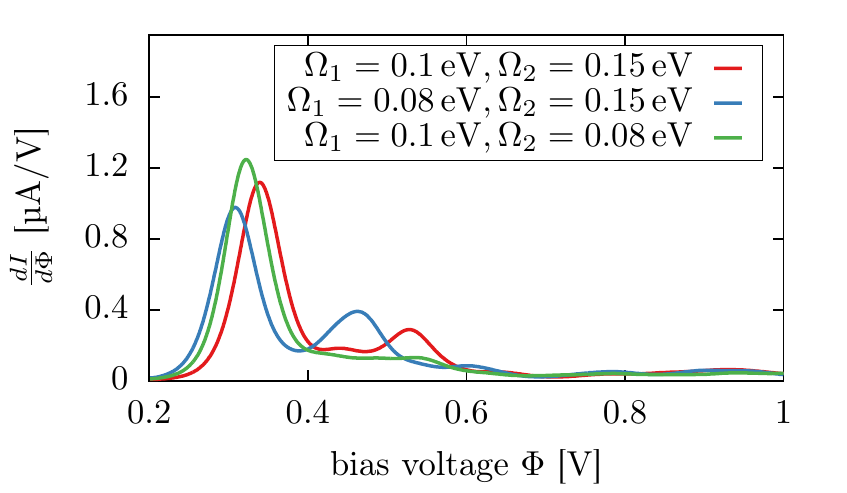}
 \caption{Conductance $\frac{d I}{d \Phi}$ as a function of bias voltage for the electronic-vibrational coupling to two vibrational modes with different vibrational frequencies.
          Furthermore, we set temperature $T=\SI{100}{\kelvin}$ and $\Gamma=2\pi V^2=\SI{1}{\milli\electronvolt}$.}
 \label{fig: VF2D}
\end{figure}

\paragraph{Dependence on electronic-vibrational coupling strength.}
To provide further insight in the transport behavior, we depict in Fig.\ \ref{fig: VF2D} the conductance as a function of bias voltage for different vibrational frequencies.

We first focus on reducing the frequency of the adiabatically coupled mode, which corresponds to the blue line in Fig.\ \ref{fig: VF2D}.
As a consequence of the accompanying increased dimensionless adiabatic coupling strength, transport processes of higher vibrationally excited states are more involved in the charge transport.
Besides, the enhanced adiabatic electronic-vibrational coupling suppresses the influence of nonadiabatic effects, since the current-voltage characteristics can be rationalized in terms of vibrationally excited states of the adiabatically coupled mode.

The effect of a reduced vibrational frequency of the nonadiabatically coupled mode is demonstrated by the green line in Fig.\ \ref{fig: VF2D}.
In the intermediate bias voltage regime between $\Phi = \SI{0.6}{\volt}$ and $\Phi =\SI{0.8}{\volt}$, i.e., when the nonadiabatic process becomes relevant, the conductance reveals a slightly different transport behavior.
An increased dimensionless nonadiabatic coupling strength leads to a more pronounced asymmetry between the two transport processes $\ket{00;00} \stackrel{\text{LS}}{\longrightarrow} \ket{\overline{10;00}}$ and $\ket{00;00} \stackrel{\text{LS}}{\longrightarrow} \ket{\overline{01;00}}$ and is not causing the onset of current suppression, as previously shown for single-mode models.

\subsubsection{Transport in a degenerate two-state two-mode model} \label{sec: DegenerateD}
Particularly interesting and well-studied examples of symmetry-induced conical intersections occur in molecules that exhibit Jahn-Teller effects \cite{koppel2009jahn,bersuker2013jahn}.
According to the Jahn-Teller theorem \cite{doi101098rspa19370142}, the molecule undergoes a geometrical distortion to form a system with lower symmetry, which has the effect of partly removing the degeneracies and reducing the overall energy.
Charge transport through model systems which exhibit Jahn-Teller effect has been studied before in Refs.\ \cite{PhysRevB77075323,PhysRevB78233401,Reckermann2008,YING201433}.

More specifically, we consider here a model of $E\otimes e$ symmetry represented in the system Hamiltonian in Eq.\ (\ref{eq: SystemHamiltonian}) by two vibrational modes with degenerate frequencies $\Omega$ and two electronic states with degenerate energies $\epsilon$.
Here, we modify the spectrum of the harmonic oscillators by adding the constant term $\Omega$ in the system Hamiltonian.
The neutral molecule, which can be described by the two-dimensional isotropic harmonic oscillator basis functions in polar representation $\ket{l,n}$ with radial excitation $n\in \mathbb{N}_0$ and angular momentum $l\in \mathbb{Z}$, has then the energy $(n+1)\Omega$.

For a model exhibiting degeneracies, coherences between degenerate states can have a major influence on the transport characteristics \cite{PhysRevB80033302}.
The HEOM approach used in this work is well suited to treat such model systems, since coherences and populations are treated on an equal footing.

\paragraph{Selection rule.}
To simplify the analysis of the transport behavior, we exploit the symmetry of the model as introduced in Refs.\ \cite{doi101098rspa19580022,PhysRev1061195,PhysRev1081251,doi1010631450519}.
We first describe the electronic manifold in a pseudospin notation with the electronic angular momentum operator $\boldsymbol{\sigma}$.
Due to the presence of the conical intersection, the corresponding eigenvalues must be half-integer for the charged molecule.
To this end, we assign the eigenvalue $s=\frac{1}{2}$ ($s=-\frac{1}{2}$) to the singly occupied electronic state $\ket{10}$($\ket{01}$), while $s=0$ holds for the neutral molecule.
Second, we introduce the vibrational angular momentum operator $\boldsymbol{l}$ whose eigenstates correspond to the two-dimensional isotropic harmonic oscillator basis functions in the polar representation $\ket{l,n}$, where $l\in \mathbb{Z}$ is the angular momentum and $n\in \mathbb{N}_0$ the radial excitation number.
Overall, we combine the electronic pseudospin basis with the two-dimensional harmonic oscillator basis to the vibronic states $\ket{s;j-s,n}$ by including the quantum number $j = l + s$ as the vibronic angular momentum.
In general, the eigenstates of the charged molecule can be expanded by
\begin{align}
 \ket{\overline{1;j,n}} = \sum_{\tilde{n},\pm j} \big( & A^{j}_{n,\tilde{n}} \ket{\frac{1}{2};j-\frac{1}{2},\tilde{n}} \nonumber \\
 &+ B^{j}_{n,\tilde{n}} \ket{-\frac{1}{2};j+\frac{1}{2},\tilde{n}} \big),
\end{align}
with real coefficients $A^{j}_{n,\tilde{n}}$ and $B^{j}_{n,\tilde{n}}$ obtained by numerical diagonalization.
Thereby, only states with the same value of $j$ are coupled.
Characteristic for the eigenenergies is the dimensionless electronic-vibrational coupling strength, which we set to $\lambda/\Omega=1$ in the following.
Moreover, we choose the frequency $\Omega$ as the general unit.
Table \ref{tab: EVJTM} in the Appendix summarizes the eigenenergies of selected energetically low-lying eigenstates in the case of $\epsilon/\Omega=0$.

The shape of the potential energy surfaces in the degenerate two-state two-mode model corresponds to the mexican hat potential \cite{doi10108000268978100101721}.
The azimuthal symmetry reflects the existence of an additional conserved quantum number, which is the vibronic angular momentum $j$.
As a consequence, some charging/decharging transitions are prohibited resulting in a selection rule \cite{PhysRevB73155306}.
We utilize the nonadiabatic transition probabilities [Eq.\ (\ref{eq: FCME})] to determine which transitions are allowed.
Specifically for the degenerate two-state two-mode model, the quantity $P^{i,-}_{s_1,j_1n_1;s_2,l_2n_2}$ with $s_1=\pm \frac{1}{2}$ and $s_2=0$ corresponds to the probability of the molecule to undergo the following transition: 
Being initially in the charged configuration $\ket{s_1;j_1 - s_1,n_1}$, an electron leaves the molecule and the final state corresponds to the neutral state with angular momentum $l_2$ and radial excitation $n_2$.
The transition probabilities vanish unless the angular momentum fulfills \cite{PhysRevB77075323}
\begin{align}
 j_1 =l_2 \pm \frac{1}{2}.
\end{align}

\begin{figure}
 \centering
 \includegraphics[width=0.5\textwidth]{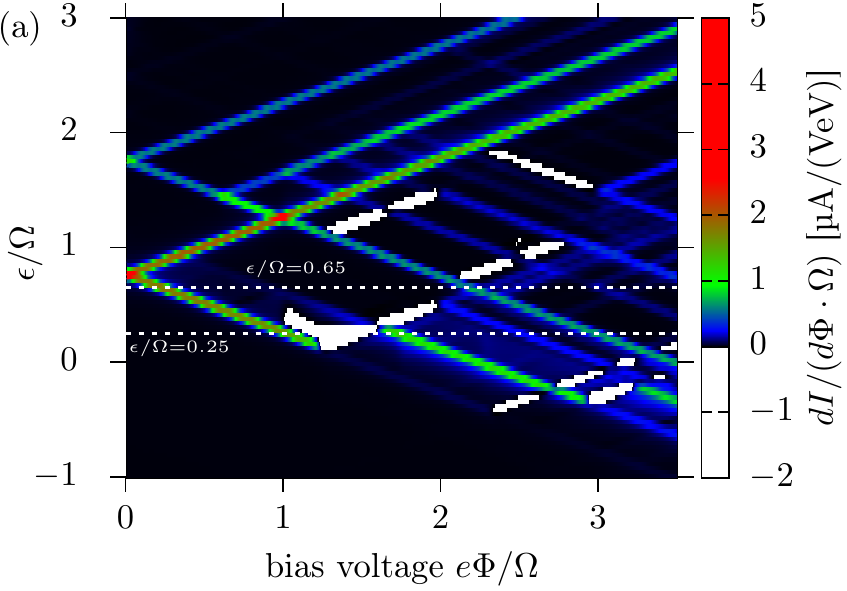}
 \includegraphics[width=0.5\textwidth]{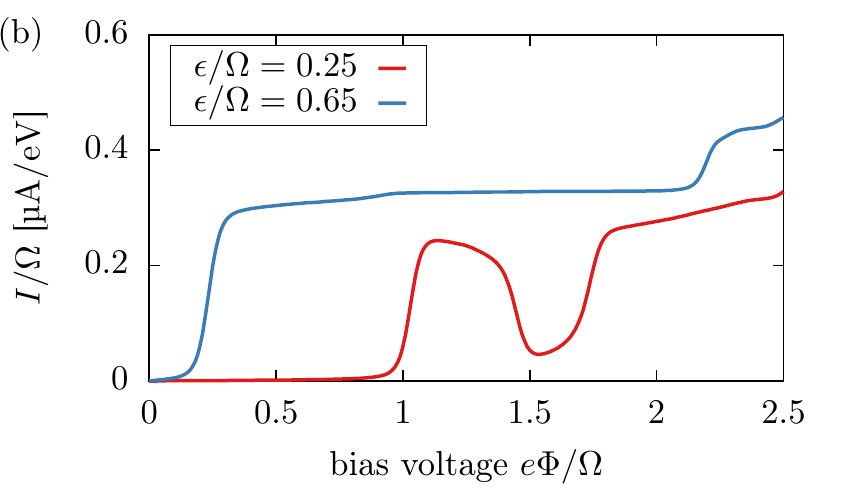}
 \caption{(a) Conductance $\frac{1}{\Omega}\frac{d I}{d \Phi}$ as a function of bias voltage and energy of the electronic levels for the two-state two-mode Jahn-Teller model.
          (b) Current-voltage characteristics for the two-state two-mode Jahn-Teller model which corresponds to a section of the conductance map in panel (a).
          All quantities are given in units of the vibrational frequencies $\Omega$.
          Furthermore, we set $\Gamma=\frac{T}{2} = 4.5 \times 10^{-3} \Omega$.
          }
 \label{fig: 2DMap}
\end{figure}

\paragraph{Transport phenomena.}
The conductance map as a function of the bias voltage and energy of the electronic levels is depicted in Fig.\ \ref{fig: 2DMap}(a).
The strongly correlated dynamics between electronic and vibrational degrees of freedom manifests itself in the large variety of transport features.
In the results, the phenomenon of NDC occurs which will be the primary focus of the following investigations.
\citeauthor{PhysRevB77075323} have investigated in Ref.\ \cite{PhysRevB77075323} the underlying transport mechanisms with a rate equation approach.
As shown below, there are differences to the results of the HEOM method used in this work, which motivate further investigations.

We show in Fig.\ \ref{fig: 2DMap}(b) the current-voltage characteristics for the selected energies $\epsilon/\Omega=0.25$ and $\epsilon/\Omega=0.65$.
The two cases are representative for two different transport regimes.
While the results for the larger electronic energy do not show any irregularities, the model parameters with the lower electronic energy lead to NDC, which we study in the following.

The first step in the current for $\epsilon/\Omega=0.25$ marks the onset of resonant tunneling through the Jahn-Teller eigenstate $\ket{\overline{1;\frac{1}{2},0}}$, which corresponds to the vibrational ground state of the charged molecule.
Further increasing the bias voltage leads to the reduction of the current which can be explained as follows.
At bias voltage $e\Phi/\Omega\approx 1.47$, the eigenstate $\ket{\overline{1;\frac{1}{2},1}}$ enters the bias window and contributes as an additional transport channel.
The decharging process $\ket{\overline{1;\frac{1}{2},1}} \stackrel{\text{SR}}{\longrightarrow} \ket{00;11}$ can then increase the radial excitation and subsequently, an electron transferred from the left lead to the molecule can cause a transition to the eigenstate $\ket{\overline{1;\frac{3}{2},0}}$.
Once the molecule is in this Jahn-Teller eigenstate, the only energetically allowed transition to a neutral state, which corresponds to the process $\ket{\overline{1;\frac{3}{2},0}} \stackrel{\text{SR}}{\longrightarrow} \ket{00;00}$, is prohibited due to the conservation of vibronic angular momentum.
Allowed transitions to other neutral states, e.g., $\ket{00;22}$, are energetically suppressed, since the energy of eigenstate $\ket{\overline{1;\frac{3}{2},0}}$ is considerably lower than the energy of the neutral vibrationally excited states $(n+1)\Omega$ with $n \ge 1$.
Overall, the combination of the special spectrum with the selection rule leads to trapping states.

In the degenerate two-state two-mode model, two mechanisms provide escape routes from trapping states.
First, higher-order processes enable energetically suppressed transitions, as discussed below.
Second, increasing the bias voltage provides enough energy to allow transport processes which are energetically suppressed at lower voltages.
In principle, every state $\ket{\overline{1;j,0}}$ with $|j|\ge\frac{3}{2}$ corresponds to a trapping state and the trapping mechanism always involves the interplay between the selection rule and the complex spectrum.

Increasing the electronic energy to $\epsilon/\Omega=0.65$ leads to the disappearance of NDC, as seen in Fig.\ \ref{fig: 2DMap}(b).
If the system is in the eigenstate $\ket{\overline{1;\frac{3}{2},0}}$, a transition to the neutral state $\ket{00;11} $ is energetically possible, since the eigenenergy of eigenstate $\ket{\overline{1;\frac{3}{2},0}}$ increases to $1.494$ and thus, the eigenstate is no longer a trapping state.

\paragraph{Higher-Order processes as escape mechanism from trapping states.}
Once the charged molecule is in a state whose transitions to neutral states are either prohibited by the selection rule or energetically suppressed, higher-order tunneling processes such as cotunneling \cite{PhysRevLett95146806,PhysRevB74205438,PhysRevB94201407,PhysRevB75205413} provide escape routes.
In previous studies \cite{PhysRevB77075323}, \citeauthor{PhysRevB77075323} employed rate equations \cite{breuer2002theory,PhysRevB77195416} to investigate charge transport through the degenerate two-state two-mode Jahn-Teller model.
Conventional rate equation approaches do not include higher-order tunneling processes and thus, exclude escape mechanisms from trapping states.
Depending on the initial conditions, the system may then occupy different trapping states resulting in different steady-state reduced density matrices \cite{PhysRevB77075323}.
Higher-order processes prevent this unphysical behavior and ensure the uniqueness of the steady-state.
\citeauthor{PhysRevB77075323} included in their rate equation formalism vibrational relaxation processes which drive the vibrational modes toward the thermal equilibrium distribution.
It thus enables mechanisms which can release the system from trapping states.
In the HEOM method used here, higher-order processes are fully included and in the following, we discuss the resulting escape processes.

\begin{figure}
\centering
 \includegraphics[width=0.5\textwidth]{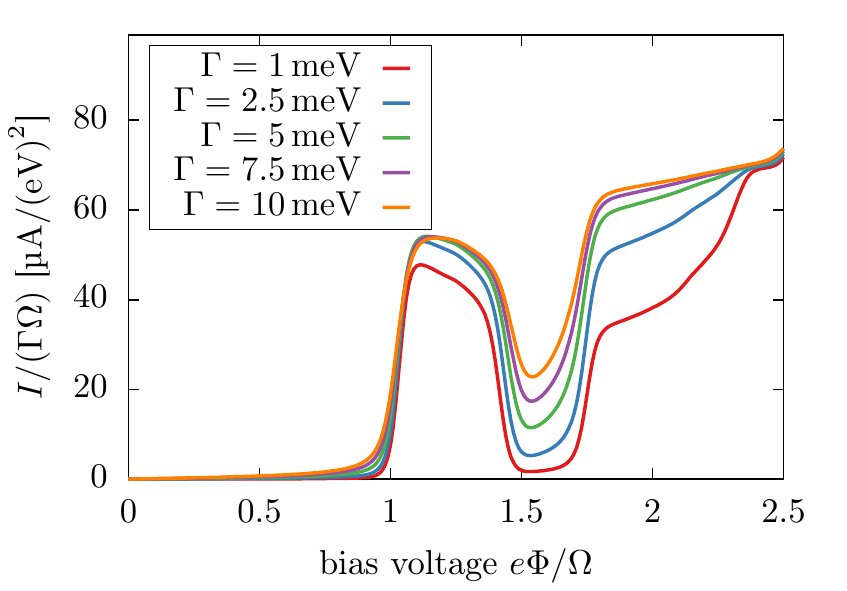}
 \caption{Current-voltage characteristics for the two-state two-mode Jahn-Teller model for different molecule-lead coupling strengths. 
          The energy of the electronic levels is set to $\epsilon/\Omega=0.25$ and the temperature to $T = 9 \times 10^{-3} \Omega$.}
 \label{fig: Cot}
\end{figure}

The transport dynamics of the Jahn-Teller model is determined by the time, during which the molecule is locked in trapping states \cite{PhysRevB77075323}, as well as the timescale of cotunneling processes, which is in the order of $1/\Gamma^2$.
In the following, we investigate the influence of the molecule-lead coupling strength on the transport characteristics affected by trapping states.
To this end, we depict in Fig.\ \ref{fig: Cot} the current-voltage characteristics for different molecule-lead coupling strengths.
We have chosen the energy of the molecular electronic level to be $\epsilon/\Omega=0.25$, where the system gets locked in Jahn-Teller eigenstate $\ket{\overline{1;\frac{3}{2},0}}$.
The nonzero current at bias voltage $e\Phi/\Omega\approx 1.55$ highlights the importance of cotunneling processes as escape routes from trapping states.
The minimum current at the trapping state increases along with the molecule-lead coupling, as explained in the following.

Cotunneling generally involves the virtual occupation of states by energetically prohibited transitions.
It becomes the dominant contribution to the current when the charged system is in a trapping state.
In general, cotunneling can be classified into two categories involving inelastic and elastic processes.
Considering the latter, the system remains in the same molecular state as it was before the tunneling event.
If the molecule is locked in a trapping state, then elastic cotunneling does not provide an escape mechanism.
Contrarily, inelastic cotunneling changes the state of the system and thus provides processes which enable the escape from trapping states by allowing energetically suppressed transitions.
To be specific, the previously discussed trapping state $\ket{\overline{1;\frac{3}{2},0}}$ can be released by the decharging process $\ket{\overline{1;\frac{3}{2},0}} \stackrel{\text{SR}}{\longrightarrow} \ket{00;11} $, followed by a repeated charging of the molecule.
In general, increasing the molecule-lead coupling strength results in a reduced timescale of the cotunneling processes.
Accordingly, the time which the molecule spends in trapping states is decreased and leads to an enhanced current through the system, as seen in Fig.\ \ref{fig: Cot}.

\section{Conclusion} \label{sec: Conclusion}
In this paper, we have employed the numerically exact HEOM approach to study nonadiabatic electronic-vibrational effects in nonequilibrium charge transport through single-molecule junctions.
More specifically, we have considered a molecular junction described by a model with two electronic levels and investigated the transport behavior for an increasing number of vibrational degrees of freedom.

In case of a single vibrational mode, we first studied the dependence of the transport properties on the nonadiabatic electronic-vibrational coupling strength.
Intermediate couplings entail the onset of low-bias current suppression which is based on the suppression of transition between energetically low-lying vibrational states similar to the Franck-Condon blockade reported in adiabatic polaron-type transport.
In the regime of strong electronic-vibrational coupling, nonadiabatic effects induce a mechanism which, depending on the applied bias voltage, effectively decouples one electronic level from the charge transport eventually resulting in the phenomenon of negative differential conductance.
Investigating the interplay between adiabatic and nonadiabatic electronic-vibrational coupling, our numerical results revealed a strong low-bias current suppression already for low to intermediate electronic-vibrational coupling strengths.

In addition, we investigated two-mode models whose transport characteristics depend profoundly on the energy spacing of the electronic states.
For a model with well-separated electronic states, we showed that the transport behavior is dominated by adiabatic polaron-type transport processes in the low-bias voltage regime while nonadiabatic transport mechanism play an enhanced role for intermediate bias voltages.
As an example for molecules with degenerate electronic states, we investigated a model exhibiting Jahn-Teller effects.
Extending previous studies \cite{PhysRevB77075323}, we found that the molecule can become trapped in a nonconducting state leading to a current-blockade.
Higher-order processes such as inelastic cotunneling, which are fully included in the HEOM approach, provide an escape mechanism from the trapping states. 
Our results revealed that an increasing molecule-lead coupling strength can partly lift the current-blockade.

In the present work, we have focused on weak to intermediate electronic-vibrational coupling strengths in the scenario of two vibrational modes.
To investigate the strong coupling regime, the treatment of the vibrational modes as a part of the reservoir subspace \cite{PhysRevB97235429,PhysRevB103235413,C6FD00088F,doi101021acsjpclett5b02793} is an interesting topic for future research.

\section*{Acknowledgements}
We thank S.\ Rudge for helpful discussions.
This work was supported by a research grant of the German Research Foundation (DFG). 
Furthermore, support from the state of Baden-{W\"urttemberg} through bwHPC and the DFG through Grants No.\ INST 40/575-1 FUGG (JUSTUS 2 cluster) and No.\ INST 37/935-
1 FUGG (bwForCluster BinAC) is gratefully acknowledged.
A.\ E.\ was supported by the DFG (Grant No.\ 453644843).

\appendix
\section{Eigenenergies of the isolated molecule}
In this Appendix, we list the eigenenergies of selected eigenstates of the isolated molecule. 
\begin{table}[H]
 \begin{center}
 \begin{tabularx}{\columnwidth}{lXXXXXXX}
 \hline
 \hline
 Model & $\Omega_1$ & $\ket{\overline{10;0}}$ & $\ket{\overline{10;1}}$ & $\ket{\overline{10;2}}$ & $\ket{\overline{01;0}}$ & $\ket{\overline{01;1}}$ & $\ket{\overline{11;0}}$ \\
 \hline
 NONAD  & $50$  & $184$  & $185$ & $225$ & $414$ & $511$ & $650$ \\
        & $60$  & $199$  & $206$ & $247$ & $418$ & $530$ & $650$ \\
        & $80$  & $213$  & $234$ & $289$ & $463$ & $564$ & $650$ \\
        & $150$ & $227$  & $299$ & $420$ & $449$ & $612$ & $650$ \\
        & $200$ & $231$  & $481$ & $503$ & $330$ & $703$ & $650$ \\
        & $220$ & $232$  & $494$ & $533$ & $338$ & $737$ & $650$ \\
        & $240$ & $233$  & $509$ & $562$ & $346$ & $770$ & $650$ \\
 INTPLY & $110$ & $78.9$ & $276$ &       & $571$ & $183$ & $417$ \\
        & $120$ & $96.9$ & $301$ &       & $553$ & $209$ & $437$ \\
        & $130$ & $112$  & $320$ &       & $538$ & $231$ & $453$ \\
        & $150$ & $135$  & $265$ &       & $443$ & $482$ & $479$ \\
        & $200$ & $171$  & $315$ &       & $413$ & $508$ & $522$ \\
        & $220$ & $180$  & $327$ &       & $436$ & $535$ & $534$ \\
        & $240$ & $187$  & $335$ &       & $458$ & $560$ & $543$ \\
        \hline
        \hline
 \end{tabularx}
\end{center}
 \caption{Eigenenergies of selected energetically low-lying eigenstates of the isolated molecule for the investigated models with a single vibrational mode. All energies are given in $\SI{}{\milli\electronvolt}$.}
 \label{tab: EVSingle}
\end{table}
\begin{table}[H]
 \begin{center}
 \begin{tabularx}{\columnwidth}{XXXXXX}
 \hline
 \hline
 $\Omega_1$ & $\Omega_2$ & $\ket{\overline{10;00}}$ & $\ket{\overline{10;10}}$ & $\ket{\overline{10;01}}$ & $\ket{\overline{01;00}}$ \\
 \hline
 $100$ & $150$ & $172$ & $268$ & $273$ & $319$ \\
 $100$ & $80$  & $167$ & $207$ & $247$ & $311$ \\
 $80$  & $150$ & $158$ & $236$ & $267$ & $314$ \\
 \hline
 \hline
 \end{tabularx}
\end{center}
 \caption{Models with two vibrational modes and well-separated electronic states: eigenenergies of selected energetically low-lying eigenstates of the isolated molecule computed for different vibrational frequencies. Depending on the electronic-vibrational coupling strength, we consider thereby up to $N_{\textrm{V},1/2}=9$ vibrational basis states. All energies are given in $\SI{}{\milli\electronvolt}$.
}
 \label{tab: EVTNonD}
\end{table}
\begin{table}[H]
 \begin{center}
 \begin{tabularx}{\columnwidth}{XXXX}
 \hline
 \hline
 $\ket{\overline{1;\frac{1}{2},0}}$ & $\ket{\overline{1;\frac{1}{2},1}}$ & $\ket{\overline{1;\frac{3}{2},0}}$ & $\ket{\overline{1;\frac{5}{2},0}}$ \\
 \hline
 $0.233$  & $1.488$ & $0.844$ & $1.540$ \\
 \hline
 \hline
 \end{tabularx}
\end{center}
 \caption{Vibronic eigenenergies of selected energetically low-lying eigenstates of the isolated molecule computed for the degenerate two-state two-mode model.
          The energy of the electronic levels was set to $\epsilon/\Omega=0$.}
 \label{tab: EVJTM}
\end{table}

\bibliography{myBib}
\end{document}